\documentclass{epl}

\title{Entanglement storage in atomic ensembles}
\author{A. Dantan, A. Bramati \and M. Pinard}
\institute{Laboratoire Kastler Brossel, Universit\'{e} Pierre et
Marie Curie,\\
4 place Jussieu, 75252 Paris Cedex 05, France}
\pacs{03.67.-a}{Quantum information} \pacs{03.67.Mn}{Entanglement
production, characterization and manipulation}
\pacs{42.50.Lc}{Quantum fluctuations, quantum noise and quantum
jumps}

\begin{document}

\maketitle

\begin{abstract}
We propose to entangle macroscopic atomic ensembles in cavity
using EPR-correlated beams. We show how the field entanglement can
be almost perfectly mapped onto the long-lived atomic spins
associated with the ground states of the ensembles, and how it can
be retrieved in the fields exiting the cavities after a variable
storage time. Such a continuous variable quantum memory is of
interest for manipulating entanglement in quantum networks.
\end{abstract}

Entanglement is one of the most intriguing feature of quantum
mechanics and, since the enunciation of the famous
Einstein-Podolsky-Rosen paradox \cite{einstein}, has always
attracted a lot of attention. In particular, it is at the heart of
quantum communication and quantum information protocols such as
quantum cryptography, teleportation, dense coding, quantum
computing \cite{bennett}. The past few years have seen many
realizations of entangled beams in the continuous variable regime,
using $\chi^{(2)}$ process in optical parametric amplifiers (OPAs)
\cite{ou,zhang,bowen,laurat}, Kerr effect in optical fibers
\cite{silberhorn,glockl} or in cold atoms \cite{josse2}. Efficient
sources of entangled beams now exist and strong correlations have
been achieved over rather broad bandwidth \cite{laurat}. In order
to build quantum communication networks in which light beams
connect atomic ensembles, a major issue is to be able to store
entanglement into the atoms \cite{polzik,lukin}. Entanglement
between two atomic ensembles has been successfully demonstrated by
Julsgaard \textit{et al.} by sending pulses of coherent light
through two atomic vapor cells \cite{julsgaard} and measuring the
outgoing field. However, the possibility to store entanglement
between quantum-correlated beams into atoms remains to be
demonstrated. In this Letter we propose a cw scheme to achieve
entanglement between two cold atom ensembles placed in cavities by
using EPR-entangled beams, as produced by OPAs for instance, and
coherent control fields. The entanglement between the beams is
mapped onto the ground state spins of the atoms and no measurement
of the field is required. Given the long lifetime of the cold
atoms spin the entanglement can thus be stored for a rather long
time when the control field is switched off. It can then be
retrieved in the vacuum modes exiting the cavities by switching on
the control field again after a variable storage time. We then
give a method to directly measure the entanglement of the outgoing
beams - and, consequently, the atomic entanglement - in one
simultaneous measurement of the EPR variances with two homodyne
detections and a single local oscillator.\\

\begin{figure} \onefigure[width=14.25cm]{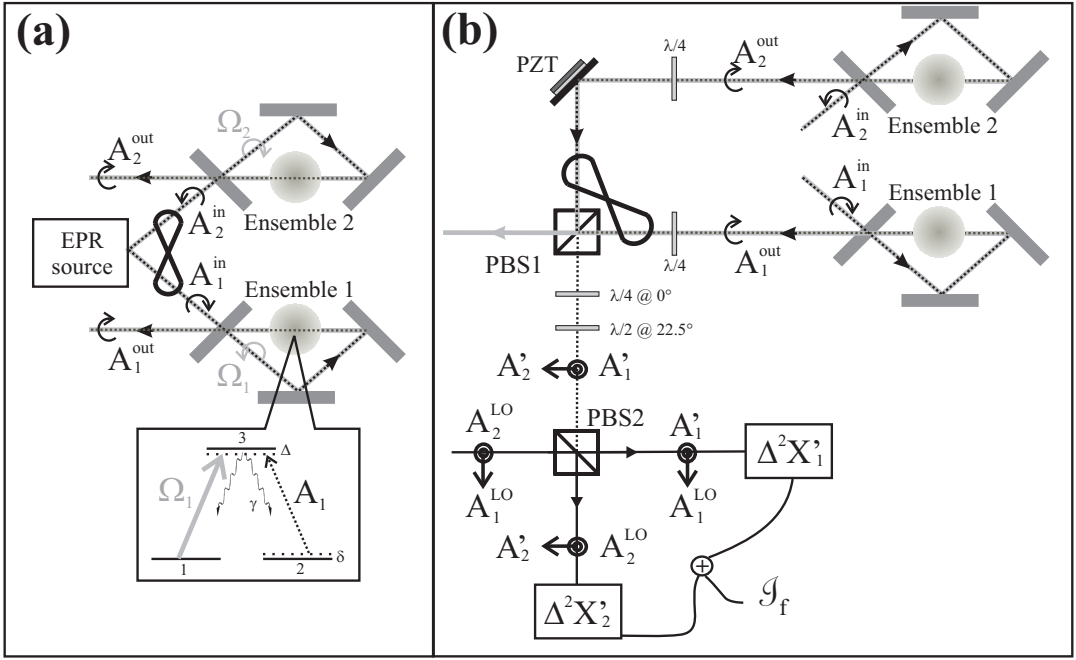} \caption{(a)
Entanglement mapping onto the atoms: the incoming vacuum fields
$A_{1,2}^{in}$ are entangled, $\Omega_{1,2}$ are intracavity
coherent control fields. The insert shows the $\Lambda$ structure
of the atomic levels considered. (b) Readout: the incident vacuum
fields are in a coherent vacuum state, the outgoing vacuum fields
are entangled when the control fields are switched on again. PBS:
polarizing beamsplitter, LO: local oscillator, PZT: piezo-electric
ceramic.} \label{setup}
\end{figure}

As shown in fig.~\ref{setup}, we consider two identical sets of
$N$ $\Lambda$-type 3-level atoms, each set interacting with a
control field $\Omega_i$, ($i=1,2$) and with one of the
EPR-entangled beams $A_i$. Without loss of generality we assume
the control fields to be $\sigma_+$-polarized and the entangled
vacuum fields $\sigma_-$-polarized. The entangled vacuum fields
can be obtained for instance from OPAs \cite{bowen} or from an OPO
below threshold \cite{laurat}, although other schemes can be
equivalently envisaged. To simplify, the entanglement bandwidth is
assumed to be larger than the cavity bandwidth $\kappa$.
 First, we examine how to create entanglement between
the atoms and start by studying the interaction of light with one
ensemble. In previous works \cite{dantan1,dantan3,dantan2}, we
showed that "Raman"- or "EIT"-type interaction of light with
$\Lambda$-type atoms could lead to squeeze the atomic spin, either
in a non linear regime, when the incoming field is in a coherent
state ("self spin squeezing"), or by transfer, when the incoming
field is a broadband squeezed vacuum. In the latter, we examined
how squeezing could be transferred from the field to the atoms in
different configurations. A quasiperfect transfer is predicted
either in an "EIT" (on one- and two-photon resonance) or a "Raman"
(on two-photon resonance, but large one-photon detunings)
configurations. Basing ourselves on these results we consider for
instance an EIT situation in which the fields are both one- and
two-photon resonant ($\Delta=\delta=0$) and cancel the cavity
detuning. Since $\langle A_1^{in}\rangle=0$, the atoms are pumped
by control field $\Omega_1$ into level $|2\rangle$ and the spin is
aligned along $z$ in steady state: $\langle J_{z1}\rangle=N/2$.
The collective atomic spin can then be treated as a harmonic
oscillator, the non-commutating spin components $J_{x1}$ and
$J_{y1}$ in the plane orthogonal to the mean spin playing the same
role as the field quadrature operators $X_1=A_1+A_1^{\dagger}$ and
$Y_1=i(A_1^{\dagger}-A_1)$. We thus seek to map the EPR
correlations existing between the quadratures $X_i$, $Y_i$ onto
the two ensembles spin components $J_{xi}$, $J_{yi}$. It is
possible to choose the control field pumping rate
$\Gamma_E=\Omega_1^2/\gamma$ so that the ground state observables
evolve slowly with respect to the fields or the optical dipoles
and get simplified equations for the atomic spin fluctuations in
the Fourier domain \cite{dantan2} \begin{eqnarray} \label{spin1}
(\tilde{\gamma}_0-i\omega)\delta
J_{x1}(\omega) &=&-\beta \delta X_1^{in}(\omega)+\tilde{f}_{x1}\\
(\tilde{\gamma}_0-i\omega)\delta J_{y1}(\omega) &=&-\beta \delta
Y_1^{in}(\omega)+\tilde{f}_{y1}\label{spin2}
\end{eqnarray} where $\tilde{\gamma}_0=\gamma_0+\Gamma_E/(1+2C)$ is
the effective atomic decay constant, satisfying
$\gamma_0\ll\tilde{\gamma}_0\ll\gamma,\kappa$, $C=g^2N/T\gamma$ is
the standard cooperativity parameter, $T$ the coupling mirror
transmission, $g$ the atom-field coupling constant and
$\tilde{f}_{x1},\tilde{f}_{y1}$ are Langevin operators, the
correlation functions of which can be calculated via the quantum
regression theorem. These operators account for the noise due to
loss of coherence in the ground state ($\propto\gamma_0$) and for
the noise contribution of the optical dipole via spontaneous
emission ($\propto\Gamma_E$).
$\beta=gN\Omega_1/\gamma\sqrt{T}(1+2C)$ represents the effective
coupling with the incoming EPR field $A_1^{in}$. To simplify we
assume a symmetrical configuration for ensemble 2
($\Omega_2=\Omega_1$, same number of atoms, $\langle
J_{z2}\rangle=N/2$, etc), so that the equations for the ground
state spin fluctuations are similar to
eqs.~(\ref{spin1}-\ref{spin2}) by substituting subscript 1 by 2.
We then obtain very simple equations for the fluctuations of the
ground state spins operators
\begin{eqnarray} (\tilde{\gamma}_0-i\omega)(\delta
J_{x1}-\delta J_{x2}) &=&-\beta (\delta X_1^{in}-\delta X_2^{in})+
\tilde{f}_{x1}-\tilde{f}_{x2}\label{jx}\\
(\tilde{\gamma}_0-i\omega)(\delta J_{y1}+\delta J_{y2}) &=&-\beta
(\delta Y_1^{in}+\delta
Y_2^{in})+\tilde{f}_{y1}+\tilde{f}_{y2}\label{jy}
\end{eqnarray} which are valid if the effective optical pumping
rate satisfy $\gamma_0\ll\tilde{\gamma}_0\ll\kappa,\gamma$. Since
the fluctuations of both spins are related to those of the
incident beams quadratures, one expects that the field
correlations will reflect on the atoms.\\

\begin{figure}
\onefigure[width=14cm]{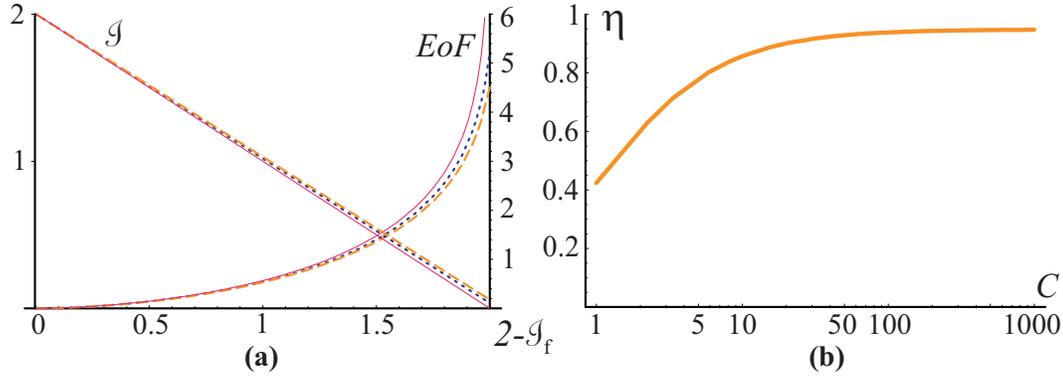} \caption{(a) Left axis:
Inseparability criterion for the incident fields $\mathcal{I}_f$
[plain] and the atoms $\mathcal{I}_{at}$ [dotted: simple model
approximation, dashed: full calculations] vs EPR correlations
$2-\mathcal{I}_f$. Right axis: entanglement of formation for
incident fields and atoms vs EPR correlations [parameters:
$C=100$, $\kappa=2\gamma$, $\gamma=1000\gamma_0$,
$\Gamma_E=15\gamma$]. (b) Optimized mapping fidelity
$\eta=f(\mathcal{I}_{at})/f(\mathcal{I}_f)$ vs cooperativity. For
each value of $C$ the pumping rate is optimized
[$\gamma=1000\gamma_0,\kappa=2\gamma,\mathcal{I}_f=1$].}
\label{eof}
\end{figure}

To quantify the entanglement we make use of the inseparability
criterion derived by Duan \textit{et al.} and Simon \cite{duan},
which rely on the sum of the variances of EPR-type Gaussian
operators, such as $X_1-X_2$ and $Y_1+Y_2$ for the field.
\begin{equation}\label{duan}
\mathcal{I}_{f}=\frac{1}{2}[\Delta^2(X_1-X_2)+\Delta^2(Y_1+Y_2)]<2\end{equation}
is a necessary condition for modes 1 and 2 to be entangled.
Moreover, when modes 1 and 2 are symmetric it has been shown by
Giedke \textit{et al.} that this condition is also sufficient and
directly related to the Entanglement of Formation (EoF), thus
providing a good measure of entanglement \cite{giedke}. For OPAs
pumped below threshold strong amplitude correlations and phase
anti-correlations exist, and both $\Delta^2(X_1-X_2)$ and
$\Delta^2(Y_1+Y_2)$ can be strongly reduced below the separable
beams value of 2 on a broad bandwidth \cite{laurat}. In a similar
fashion the EPR-type operators for the atomic ensembles are
$J_{x1}-J_{x2}$ and $J_{y1}+J_{y2}$, which satisfy $\langle
[J_{x1}-J_{x2},J_{y1}+J_{y2}]\rangle=i\langle J_{z1}-J_{
z2}\rangle=0$. The atomic ensembles are then entangled if the
following inseparability criterion is satisfied \cite{julsgaard}
\begin{equation}\label{critere}
\Delta^2(J_{x1}-J_{x2})+\Delta^2(J_{y1}+J_{y2})<|\langle
J_{z1}\rangle|+|\langle J_{z2}\rangle|=N\end{equation}
 It is then
obvious from eqs.~(\ref{jx}-\ref{jy}) that any entanglement
between modes 1 and 2 will be transferred to the atomic spin,
provided the coupling $\beta$ is large enough with respect to the
noises of the process $\tilde{f}_{\alpha i}$. This is indeed
possible in EIT or Raman situations for which the noises are
substantially reduced by the cooperative behavior of the atoms
\cite{dantan2}, while the coupling is enhanced by the number of
atoms, as we will show further. In other words, the
signal-to-noise ratio of the transfer process is increased by the
atoms cooperative behavior, even though the strong coupling regime
is not reached \cite{zoller}. Assuming broad bandwidth amplitude
correlations and phase anti-correlations and in the regime
$\gamma_0\ll\tilde{\gamma}_0\ll\gamma,\kappa$, the sum of the
atomic EPR variances,
$\mathcal{I}_{at}=\frac{2}{N}[\Delta^2(J_{x1}-J_{x2})+\Delta^2(J_{y1}+J_{y2})]$
(normalized to 2) is directly related to the amount of EPR-type
correlations of the incident beams $\mathcal{I}_{f}$
\begin{equation}\label{simple}\mathcal{I}_{at}=\frac{2C}{1+2C}\frac{\Gamma_E}{(1+2C)\tilde{\gamma}_0}
\;\mathcal{I}_{f}+2\left[\frac{\gamma_0}{\tilde{\gamma}_0}+\frac{\Gamma_E}{(1+2C)^2\tilde{\gamma}_0}\right]\end{equation}
The first term in (\ref{simple}), proportional to
$\mathcal{I}_{f}$, can be understood as the atom-field coupling
factor and can be very close to 1 for $C$ and $\Gamma_E$ large
enough. The second and third terms represent the noises of the
transfer process due to the loss of coherence in the ground state
($\propto\gamma_0$) and spontaneous emission ($\propto \Gamma_E$),
respectively. Both can be made small in the regime chosen for the
pumping and for large values of $C$. Typical experimental values
of $C$ of 100-1000 ensure a quasiperfect entanglement mapping from
fields to atoms. Note also that, if the incident fields are not
entangled ($\mathcal{I}_f=2$), so are the atoms
($\mathcal{I}_{at}=2$). The accuracy of our simple picture has
been checked by full calculations involving the three-level atomic
structure and the exact covariance matrix of the whole atom-field
system. The results are represented in fig.~\ref{eof}(a) in which
we plot $\mathcal{I}_{at}$ as a function of the EPR correlations
$2-\mathcal{I}_{f}$. Very good agreement is found between the
simple model [Eq.~(\ref{simple}), dotted] and the numerical
simulations [dashed]. As a comparison the inseparability criterion
for the incident fields $\mathcal{I}_f$ is also represented
[plain]. In a symmetrical configuration the inseparability
criterion (\ref{duan}) is a measure of entanglement via the
entanglement of formation \cite{giedke}. The incident EPR fields
and the atomic EoFs (see \cite{eof} for details) are also both
plotted in fig.~\ref{eof}(a). To quantify the efficiency of the
mapping we plot in fig.~\ref{eof}(b) the ratio of the atomic EoF
to that of the EPR fields,
$\eta=f(\mathcal{I}_{at})/f(\mathcal{I}_f)$, as a function of the
cooperativity. An excellent mapping ($\eta\sim 100\%$) is obtained
for easily accessible values of $C$.
\\

Once the field entanglement has been mapped onto the atoms the
fields can be switched off and the atoms return to the coherent
spin state very slowly, on a time scale given by $1/\gamma_0$. We
now address the readout problem and show that it is possible to
retrieve the entanglement in the vacuum fields exiting the
cavities by switching on again \textit{only} the control fields.
Indeed, starting now with correlated atomic ensembles and incoming
coherent vacuum fields we expect the fluctuations of
$J_{x1}-J_{x2}$ and $J_{y1}+J_{y2}$ to imprint on the fluctuations
of, respectively, $X_1^{out}-X_2^{out}$ and $Y_1^{out}+Y_2^{out}$.
As represented in fig.~\ref{setup} the outgoing vacuum fields are
combined using quarter-wave plates on the first PBS while the
outgoing control fields are discarded. After PBS1 one disposes of
two entangled vacuum modes with orthogonal polarization. The path
dephasing can easily be cancelled using the control fields
interference signal to lock the piezo-electric ceramic. Using the
method developed in ref.~\cite{josse3} we then rotate the
polarization basis so as to retrieve two squeezed modes for the
same quadrature. This can be easily done with a quarter-wave plate
at $0^{\circ}$ in order to rotate the noise ellipsoid of one mode
by $\pi/2$ with respect to the other. One then uses a half-wave
plate at $22.5^{\circ}$ to get the $\pm 45^{\circ}$ polarization
modes, which are now squeezed for the same quadratures. To
simultaneously measure the squeezing of both modes we use the
technique of ref.~\cite{josse2} and combine on PBS2 the beam to be
measured with a local oscillator polarized at $45^{\circ}$ to the
cube axes. We then perform two balanced homodyne detections, the
first of which measures the noise of
$X_1'=(X_1^{out}-X_2^{out})/\sqrt{2}$, the second,
$X_2'=(Y_1^{out}+Y_2^{out})/\sqrt{2}$. The sum of the signals
therefore gives directly the value of the inseparability criterion
(\ref{duan}).\\

More quantitatively one may express the correlation function of
$X_1'$ in the regime $\gamma_0\ll\tilde{\gamma}_0\ll\kappa,\gamma$
as functions of the atomic EPR variances at the switching time
\begin{eqnarray} \nonumber\mathcal{C}_1'(t,t')\equiv \langle\delta
X'_1(t)\delta
X'_1(t')\rangle=\delta(t-t')-\frac{4C\Gamma_E}{(1+2C)^2}\left[1-\frac{\Delta^2
(J_{x1}-J_{x2})}{N/2}\right]e^{-\tilde{\gamma}_0(t+t')}\end{eqnarray}
A similar expression holds for the correlation function of $X_2'$,
replacing $J_{x1}-J_{x2}$ by $J_{y1}+J_{y2}$. When the atoms are
in coherent spin states ($\Delta^2(J_{x1}-J_{x2})=N/2$) one
retrieves the standard $\delta$-function of a free field. We must
now specify how the homodyne detections are performed. In order to
correctly measure the squeezing of modes $A_1'$ and $A_2'$, and,
therefore, the field entanglement at the output of the cavity, we
choose for the LO a temporal profile in $e^{-\tilde{\gamma}_0t}$
which matches that of the vacuum modes (as can be seen from the
correlation functions). In this case the normalized power measured
by a Fourier-limited spectrum analyzer integrating over a time
large with respect to $1/\tilde{\gamma}_0$ is given by
\begin{eqnarray}
\nonumber P_1(t)&\equiv
&\int_{-\pi/T_0}^{\pi/T_0}\frac{d\omega}{2\pi}
\int_{t}^{t+T_0}d\tau \int_{t}^{t+T_0}d\tau'
e^{-i\omega(\tau-\tau')}E_{LO}(\tau)E_{LO}(\tau')\mathcal{C}_1'(\tau,\tau')\\
\nonumber &=&\mathcal{N}-\mathcal{S}\left[1-\frac{\Delta^2
(J_{x1}-J_{x2})}{N/2}\right]e^{-2\tilde{\gamma}_0t}
\end{eqnarray}
again with a symmetrical expression for the second homodyne
detection. $\mathcal{N}$ and $\mathcal{S}$ are integrals depending
on $T_0$, $\tilde{\gamma}_0$ and $C$. $\mathcal{N}$ represents the
shot noise (i.e. the noise level for uncorrelated atoms, and,
therefore, uncorrelated fields), so that $\mathcal{S}/\mathcal{N}$
is the signal-to-noise ratio of the measurement. $\mathcal{N}$ can
be easily measured in a preliminary experiment in which the atoms
are prepared in a coherent spin state. The signal-to-noise ratio
can be shown to be close to 1 when $\tilde{\gamma}_0T_0\gg 1$ and
$C\gg 1$ \cite{dantan2}. Note that this is possible because we
have chosen the right matching profile for the LO. For short times
the atomic entanglement is given by
\begin{eqnarray}
\nonumber
\frac{1}{\mathcal{N}}[P_1(0)+P_2(0)]\simeq\frac{2}{N}[\Delta^2(J_{x1}-J_
{x2})+\Delta^2(J_{y1}+J_{y2})]=\mathcal{I}_{at}\end{eqnarray} We
indeed retrieve the value of the inseparability criterion
(\ref{critere}) for the atomic ensembles, i.e. measuring the
atomic entanglement
is equivalent to measuring the entanglement between the outgoing modes.
It is then possible with this technique to detect the atomic entanglement
with nearly 100\% efficiency.\\

In conclusion we have proposed a scheme to achieve continuous
entanglement between atomic ensembles in cavities using a pair of
EPR-correlated beams. The entanglement can be stored for a long
time in the ground-state atomic spins and retrieved at will in the
fields exiting the cavities by switching on the control fields. We
also propose a technique to perform the atomic entanglement
readout in a single shot measurement with one local oscillator. It
is worth noticing that all the results obtained in an EIT
configuration can be readily transposed in a Raman configuration,
in which a perfect entanglement storage is also predicted in the
regime $\gamma_0\ll(1+2C)\Gamma_R\ll\kappa,\gamma$ (where
$\Gamma_R=\gamma\Omega^2/\Delta^2$ is the Raman optical pumping
rate). We would like to point out that, although the storage time
is given by the inverse of the natural decay rate of the ground
state $1/\gamma_0$, the memory bandwidth, i.e. the frequency
bandwidth over which the entanglement is stored is much broader
($\propto\tilde{\gamma}_0$), because of the cavity interaction
with the field. Such a quantum memory could allow to store and
manipulate entanglement which is a key challenge for quantum
communication and information.

\acknowledgments


\begin{thebibliography}{0}

\bibitem{einstein}
   \Name{Einstein A., Podolsky B. \and Rosen R.}
   \REVIEW{Phys. Rev.}{47}{1935}{777}.

\bibitem{bennett}
  \Name{DiVincenzo D.P.}
  \REVIEW{Science}{270}{1995}{255};
  \Name{Furusawa A., Sorensen J., Braustein S. Fuchs C., Kimble H.J. \and Polzik E.S.}
  \REVIEW{Science}{282}{1998}{706};
  \Name{Braustein S.L. \and Kimble H.J.}
  \REVIEW{Phys. Rev. A}{61}{2000}{042302};
  \Name{Li X., Pan Q., Jing J., Zhang J., Xie C. \and Peng K.}
  \REVIEW{Phys. Rev. Lett.}{88}{2002}{047904}.

\bibitem{ou}
  \Name{Ou Z.Y., Pereira S.F., Kimble H.J. \and Peng K.}
  \REVIEW{Phys. Rev. Lett.}{68}{1992}{3663}.

\bibitem{zhang}
  \Name{Zhang Y., Wang H., Li X., Jing J., Xie C. \and Peng K.}
  \REVIEW{Phys. Rev. A}{62}{2000}{023813}.

\bibitem{bowen}
  \Name{Bowen W.P., Treps N., Schnabel R. \and Lam P.K.}
  \REVIEW{Phys. Rev. Lett.}{89}{2002}{253601};
  \Name{Bowen W.P., Schnable R. Lam P.K. \and Ralph T.C.}
  \REVIEW{Phys. Rev. Lett.}{90}{2003}{043601}.

\bibitem{laurat}
  \Name{Laurat J., Coudreau T., Keller G., Treps N. \and Fabre C.}
  (2004), quant-ph/0403224.

\bibitem{silberhorn}
  \Name{Silberhorn C., Lam P.K., Wei\ss  O., K\"{o}nig F., Korolkova N. \and Leuchs G.}
  \REVIEW{Phys. Rev. Lett.}{86}{2001}{4267}.

\bibitem{glockl}
  \Name{Gl\"{o}ckl O., Lorenz S., Marquardt C., Heersink J., Brownnutt M., Silberhorn C., Pan Q., van Loock P., Korolkova N. \and Leuchs G.}
  \REVIEW{Phys. Rev. A}{68}{2003}{012319}.

\bibitem{josse2}
  \Name{Josse V., Dantan A., Bramati A., Pinard M. \and Giacobino E.}
  \REVIEW{Phys. Rev. Lett.}{92}{2004}{123601}.

\bibitem{polzik}
  \Name{Duan L.M., Cirac J.I., Zoller P. \and Polzik E.S.}
  \REVIEW{Phys. Rev. Lett.}{85}{2000}{5643}.

\bibitem{lukin}
  \Name{Lukin M.D.}
  \REVIEW{Rev. Mod. Phys.}{75}{2003}{457}.

\bibitem{julsgaard}
  \Name{Julsgaard B., Kozhekin A. \and Polzik E.S.}
  \REVIEW{Nature}{413}{2001}{400}.

\bibitem{dantan1}
  \Name{Dantan A., Pinard M., Josse V., Nayak S. \and Berman P.R.}
  \REVIEW{Phys. Rev. A}{67}{2003}{045801}.

\bibitem{dantan3}
  \Name{Dantan A., Pinard M. \and Berman P.R.}
  \REVIEW{Eur. Phys. J. D}{27}{2003}{193}.

\bibitem{dantan2}
  \Name{Dantan A. \and Pinard M.}
  \REVIEW{Phys. Rev. A}{69}{2004}{043810}.

\bibitem{duan}
  \Name{Duan L.M., Giedke G., Cirac. J.I. \and Zoller P.}
  \REVIEW{Phys. Rev. Lett.}{84}{2000}{2722};
  \Name{Simon R.}
  \REVIEW{Phys. Rev. Lett.}{84}{2000}{2726}.

\bibitem{giedke}
  \Name{Giedke G., Wolf M.M., Kr\"{u}ger O., Werner R.F. \and Cirac J.I.}
  \REVIEW{Phys. Rev. Lett.}{91}{2003}{107901}.

\bibitem{zoller}
  \Name{Duan L.M., Lukin M.D., Cirac J.I. \and Zoller P.}
  \REVIEW{Nature (London)}{414}{2001}{413}.

\bibitem{eof} Given $\mathcal{I}_{\alpha}$ for symmetric states the EoF is calculated using
$f(\mathcal{I}_{\alpha})=c_+(\mathcal{I}_{\alpha}/2)
\log_2[c_+(\mathcal{I}_{\alpha}/2)]-c_-(\mathcal{I}_{\alpha}/2)\log_2[c_-(\mathcal{I}_{\alpha}/2)]$,
where $c_{\pm}(x)=(x^{1/2}\pm x^{-1/2})^2/4$ \cite{giedke}.

\bibitem{josse3}
  \Name{Josse V., Dantan A., Bramati A. \and Giacobino E.}
  \REVIEW{J. Opt. B: Quant. Semiclass.}{}{2004}{to be published}, quant-ph/0310139.

\end{thebibliography}
\end{document}